\documentclass[%
reprint,
superscriptaddress,
amsmath,
amssymb,
aps,
prx,
floatfix,
]{revtex4-2}

\usepackage{graphicx}% Include figure files
\usepackage{dcolumn}% Align table columns on decimal point
\usepackage{bm}% bold math
\usepackage{hyperref}% add hypertext capabilities
\usepackage{xr}
\usepackage{lipsum}

\usepackage{xcolor}
\begin{document}
\preprint{APS/123-QED}
\bibliographystyle{unsrtnat} 
%=======================================================================================================================================
\title{Charge-sensing of a Ge/Si core/shell nanowire double quantum dot using a high-impedance superconducting resonator}
%%%%%
\author{J.\,H.~Ungerer}
\altaffiliation{These authors contributed equally to the work.}
%\email{jannhinnerk.ungerer@unibas.ch}
\affiliation{
Department of Physics, University of Basel, Klingelbergstrasse 82 CH-4056, Switzerland
}
\affiliation{
Swiss Nanoscience Institute, University of Basel, Klingelbergstrasse 82 CH-4056, Switzerland
}
\author{P.~Chevalier Kwon\footnotemark[1]}
\altaffiliation{These authors contributed equally to the work.}

\affiliation{
Department of Physics, University of Basel, Klingelbergstrasse 82 CH-4056, Switzerland
}
\author{T.~Patlatiuk}
\affiliation{
Department of Physics, University of Basel, Klingelbergstrasse 82 CH-4056, Switzerland
}
\author{J.~Ridderbos}
\affiliation{
Department of Physics, University of Basel, Klingelbergstrasse 82 CH-4056, Switzerland
}
\affiliation{MESA+ Institute for Nanotechnology, University of Twente, P.O. Box 217, 7500 AE Enschede, The Netherlands}
\author{A.~Kononov}
\affiliation{
Department of Physics, University of Basel, Klingelbergstrasse 82 CH-4056, Switzerland
}
\author{D.~Sarmah}
\affiliation{
Department of Physics, University of Basel, Klingelbergstrasse 82 CH-4056, Switzerland
}
\author{E.\,P.\,A.\,M.~Bakkers}
\affiliation{Kavli Institute of Nanoscience, Delft University of Technology, Lorentzweg 1, 2628 CJ Delft, The Netherlands}
\affiliation{Department of Applied Physics, TU Eindhoven, Den Dolech 2, 5612 AZ Eindhoven, The Netherlands}
\author{D.~Zumbühl}
\affiliation{
Department of Physics, University of Basel, Klingelbergstrasse 82 CH-4056, Switzerland
}
\affiliation{
Swiss Nanoscience Institute, University of Basel, Klingelbergstrasse 82 CH-4056, Switzerland
}
\author{C.~Sch{\"o}nenberger}
\homepage{www.nanoelectronics.unibas.ch}
\affiliation{
Department of Physics, University of Basel, Klingelbergstrasse 82 CH-4056, Switzerland
}
\affiliation{
Swiss Nanoscience Institute, University of Basel, Klingelbergstrasse 82 CH-4056, Switzerland
}
\date{\today}
%=======================================================================================================================================
% ABSTRACT
%=======================================================================================================================================

\begin{abstract}
Spin qubits in germanium are a promising contender for scalable quantum computers. 
Reading out of the spin and charge configuration of quantum dots formed in Ge/Si core/shell nanowires is typically performed by measuring the current through the nanowire.
Here, we demonstrate a more versatile approach on investigating the charge configuration of these quantum dots.
We employ a high-impedance, magnetic-field resilient superconducting resonator based on NbTiN and couple it to a double quantum dot in a Ge/Si nanowire.
This allows us to dispersively detect charging effects, even in the regime where the nanowire is fully pinched off and no direct current is present. 
Furthermore, by increasing the electro-chemical potential far beyond the nanowire pinch-off, we observe indications for depleting the last hole in the quantum dot by using the second quantum dot as a charge sensor. This work opens the door for dispersive readout and future spin-photon coupling in this system. 
\end{abstract}
\maketitle

%%%% INTRO %%%%

\section{Introduction}
%=======================================================================================================================================
% INTRODUCTION
%=======================================================================================================================================
The interest in group-IV semiconductor spin qubits is large because of their small footprint, a low concentration of nuclear spins and available knowledge about their production in semiconductor industry~\cite{Zwanenburg2013,Kloeffel2013,Vandersypen2017,Scappucci2021,chatterjee2021semiconductor}.
By integrating on-chip superconducting resonators, strong spin-photon coupling has been demonstrated for spins of confined electrons in a Si two-dimensional electron gas~\cite{Mi2018,Samkharadze2018}.
Hole spins may offer the additional advantages of improved relaxation and decoherence times as they lack a valley  degeneracy  and exhibit a reduced wave-function overlap with nuclear spins~\cite{Yang2013,Prechtel2016}.
Especially, holes in one-dimensional Si or Ge nanowires ~\cite{xiang2006ge,conesa2017boosting,brauns2016highly} are of a special interest because they posses strong spin-orbit interaction~\cite{kloeffel2011strong,Kloeffel2018,froning2021strong}.
The spin-orbit interaction potentially simplifies qubit control and coupling to resonators by electric-dipole spin resonance (EDSR)~\cite{Kloeffel2013a,Maier2013}.
It thereby releases the need of implementing micromagnets and hence facilitates scaling-up.
\begin{figure}[b]
\includegraphics[width=\linewidth]{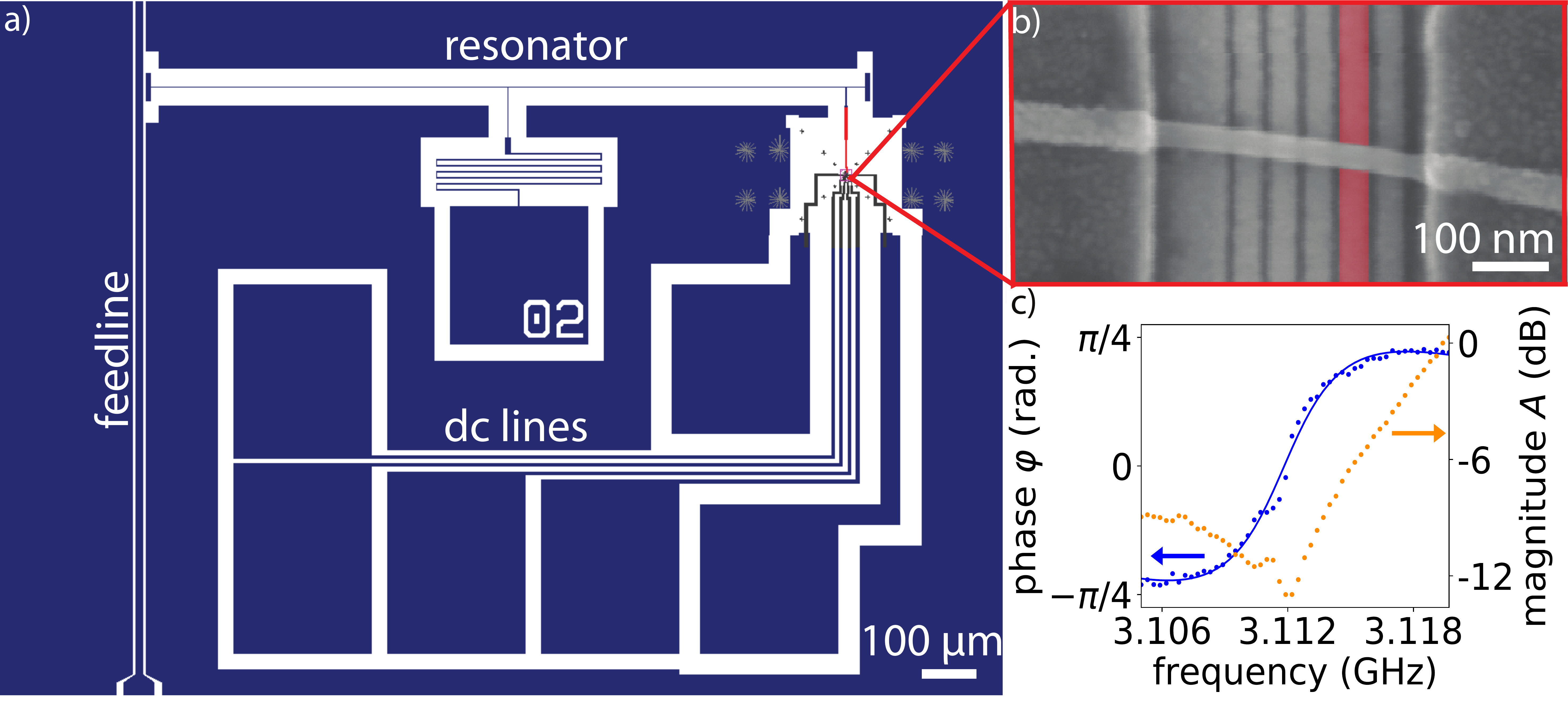}
\caption{Device overview. a) Schematic of hybrid resonator architecture. NbTiN is shown in dark blue, the Si substrate is shown white. The feedline on the left is used for reading out the notch-type coplanar-waveguide half-wave resonator which is dc biased at its voltage node in the center. Additional dc lines are used for sending current through the nanowire and applying gate voltages on all bottom gates. b)  False colored scanning electron micrograph of a similar device with Ge/Si nanowire lying on top of bottom gates covered with HfO\textsubscript{2}. The gate colored red is connected to the resonator. c) Transmission (phase and magnitude) through the feedline as a function of frequency close to the resonator frequency. The solid blue curve indicates a fit from which we extract the resonance frequency and estimate the quality factor (see main text).\label{fig:deviceoverview}}
\end{figure}

Recently, the coherent manipulation of a hole-spin qubit in a gate-defined double quantum dot (DQD) in a  Ge/Si core/shell nanowire has been demonstrated~\cite{Froning2021}.
However, in these experiments both the charge and the spin-state of the double quantum dot were determined by direct current measurements.
This technique limits the capability of determining the total number of holes present in the nanowire.
Furthermore, it requires long integration times and severely limits the maximum cycle length in pulsed-gate experiments.

Rather than measuring the current through the Ge/Si core/shell nanowire double quantum dot, pioneering works have employed another quantum dot to determine changes in the charge-occupancy of the DQD and to perform spin readout~\cite{hu2007ge,hu2012hole}.

A different approach for measuring the DQD is realized by probing a resonator coupled to the source contact of a DQD~\cite{chorley2012measuring,schroer2012radio,petersson2010charge}.
This approach is further simplified by connecting the resonator to a plunger gate, performing gate-dispersive sensing~\cite{colless2013dispersive}.
This technique has enabled measurements of the relaxation and dephasing times of hole spins in a Ge/Si core/shell nanowire DQD using a lumped-element resonator~\cite{higginbotham2014hole}.
First attempts of coupling Ge/Si nanowires to on-chip superconducting resonators were based on low-impedance resonators with a weak charge-photon coupling and in a regime of many holes present in the nanowire~\cite{wang2019gate}.

In this work, we extend the existing measurements by coupling one of the two quantum dots to a high-impedance superconducting resonator, see Fig.~\ref{fig:deviceoverview}.
The used coupling scheme allows us to detect charging in the other dot by means of capacitive charge sensing~\cite{hutin2019gate,ansaloni2020single,chanrion2020charge,borjans2021spin}.
We map the charge-stability diagram using both, direct current measurements and resonator spectroscopy.
Furthermore, we gate the nanowire to a regime of low hole occupancy where no direct current through the nanowire can be observed (pinch-off).
In this regime, the resonator spectroscopy signal reveals the presence of several more holes in the investigated dot.
Finally, by further increasing the gate voltages, we find indications of the depletion of the last hole in the investigated dot.
\section{Device description}
An overview of the device under investigation is shown in Fig.~\ref{fig:deviceoverview}a) and b). The device consists of a hybrid resonator-nanowire architecture. A notch-type half-wave ($\lambda/2$) resonator with a central frequency $f_r\approx3.1$\,GHz is defined in a NbTiN film of thickness $\sim 10$\,nm, center conductor width of $\sim 370$\,nm and a distance between center conductor and ground plane of $\sim 35$\,\textmu m. The resonator is capacitively coupled at a voltage anti-node to a feedline which is used for resonator readout. At the middle of the center conductor (voltage node), the resonator is dc biased. In front of the dc bias pad, a meandered inductor ensures sufficient frequency detuning between the $\lambda/2$ mode and a second, quarter-wave mode that forms due to the T-shaped section of the resonator. Thereby, microwave-leakage through the dc bias line is reduced~\cite{Harvey-Collard2020}. The resonator's second voltage anti-node is galvanically connected to one out of five bottom gates. The bottom gates are fabricated by Ti/Pd sandwiched by ALD-grown HfO\textsubscript{2} and have a width of approximately 25\,nm. The gate pitch is 50\,nm. On top of the bottom gates a Ge/Si core/shell nanowire is deterministically placed using a micromanipulator, see Fig.~\ref{fig:deviceoverview} b). All presented measurements are performed in a dilution refrigerator at a base temperature of 35\,mK.

The transmission $S_{21}$ through the feedline in proximity to the notch-type resonator as a function of frequency $f$ is given by~\cite{petersan1998measurement,Probst2015}
%#following equation from S.P.
%1/Ql=1/Qc+1/Qi
%-> Ql=1/(1/Qc+1/Qi)
%-> Q_l/Q_c=1/(Qc(1/Qc+1/Qi))
          %=1/(1+Qc/Qi)
%Now Qc->Qc/2
%->Ql/Qc=1/(1+Q_c/(2Q_\mathrm{loss}))
%Ql=1/(2/Q_c+1/Q_\mathrm{loss})
%\begin{equation*}\label{eq:S21}
%    \resizebox{\hsize}{!}{%
%        $S_{21}(f)=ae^{i\alpha}e^{-2\pi if\tau}\left[1-\frac{e^{i\Phi}/(1+Q_c/(2Q_\mathrm{loss}))}{1+2i(f/f_r-1)/(2/Q_c+1/Q_\mathrm{loss})}
%\right],$%      
%        }
%\end{equation*}
$$$$
\begin{equation*}
\label{eq:S21}
\resizebox{\hsize}{!}{%
        $S_{21}(f)=ae^{i\alpha}e^{-2\pi if\tau}        \left[1-\frac{e^{i\Phi}/ \left(1+Q_c/Q_\mathrm{loss}\right)}    {1+2i \left(f/f_r-1\right)
/\left(1/Q_c + 1/Q_\mathrm{loss}\right)
}\right],$
        }
\end{equation*}
where $a$, $\alpha$ and $\tau$ account for the microwave propagation through the wiring in the cryostat and the resonance is described by its resonance frequency $f_r$, the coupling quality factor $Q_\mathrm{c}$ and the loss quality factor $Q_\mathrm{loss}$. The term $e^{i\Phi}$ accounts for the Fano shape of the observed resonance arising from impedance mismatches in the feedline coupled to the resonator~\cite{Khalil2012}.

We identify the resonance of the superconducting resonator at around 3.1\,GHz by considering its temperature dependence. The measured transmission (phase and magnitude) through the feedline around resonance is shown on Fig.~\ref{fig:deviceoverview}c). The signal is superimposed on a large standing-wave background  (see Fig.~\ref{fig:extradatafig1} in the appendix.) which we attribute to an impedance mismatch between the feedline and the 50-Ohm environment of the cryostat. Despite the large fluctuations in the transmission magnitude, we are able to fit the phase of the transmission (solid, blue curve in Fig.~\ref{fig:deviceoverview}c) and extract the resonance frequency $f_r=3.111$\,GHz, and estimate the Q factors $Q_\mathrm{c}\approx600$ and $Q_\mathrm{loss}\approx600$. The uncertainity in these values originates from the large standing wave background.

We perform a finite-element simulation of the resonator using Sonnet and recover the resonance frequency of the central mode of the resonator half-wave mode when taking into account a sheet kinetic inductance of $70$\,pH/$\square$. Together with the stray line capacitance of 75 pF/m, this corresponds to a resonator impedance of 1.6\,k$\Omega$, much larger than the standard 50\,$\Omega$, hence improving the coupling strength between resonator and double quantum dot~\cite{blais2004cavity,samkharadze2016high}. We attribute the rather low $Q_\mathrm{loss}$ to microwave leakage from the resonator to the dc lines via capacitive coupling through the set of bottom gates~\cite{Mi2017}. Indeed, using Sonnet, we estimate the mutual capacitance between two neighbouring bottom gates to be $C_\mathrm{gg}\approx 800$\,aF.
In future works, the mutual capacitance can likely be decreased with an optimised gate geometry and the resulting microwave leakage might be further reduced via improved filtering of the dc lines~\cite{mi2017circuit,Harvey-Collard2020}.

%%%% CHARGE SENSING %%%%

\section{Charge sensing}
\begin{figure}[!h]
\includegraphics[width=\linewidth]{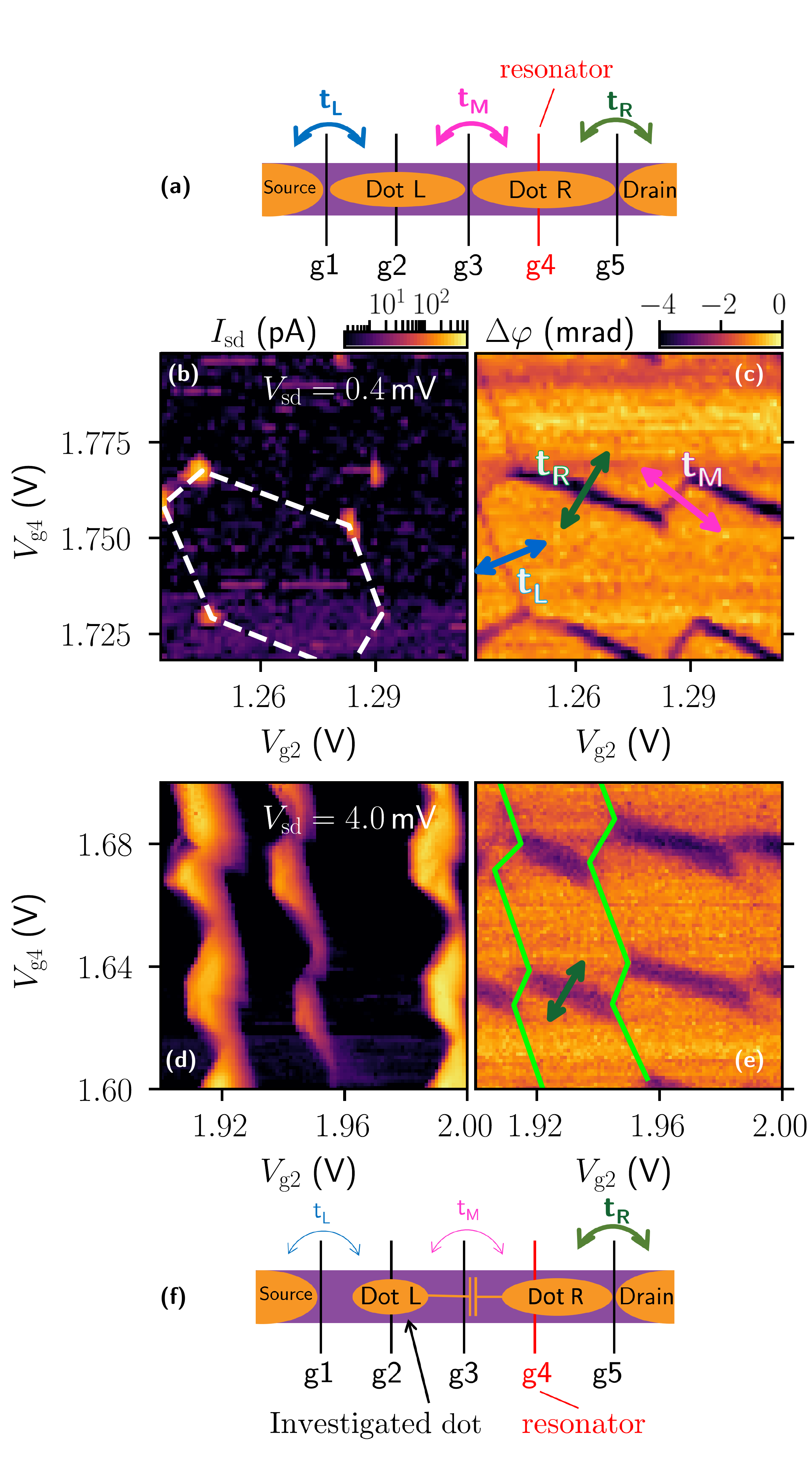}
\caption{Charge sensing. a) Schematic of the gate-defined double quantum dot and the relevant tunnel couplings between dots and leads. b) Logarithmic current, $I_\mathrm{sd}$, through the nanowire exhibiting the position of triple points. Here, the bias voltage is $V_\mathrm{sd}=400$\,\textmu V and the values of the other gate voltages are $V_\mathrm{g1}=3.2$\,V, $V_\mathrm{g3}=1.175$\,V, $V_\mathrm{g5}=9.0$\,V c) Phase difference, $\Delta \varphi$ of the resonator spectroscopy acquired at the same time as b). Tunnel couplings depicted in a) cause a phase shift of the resonator when any potentials of the system are aligned, as indicated by the colored double arrows corresponding to the tunnel transitions in a). f) Schematic of double quantum dot for a more isolated configuration. d) and e) correspond to b) and c) for the configuration depicted in f). Solid, green lines in e) indicate discharging lines of dot L.
Here, the values of the other gate voltages are $V_\mathrm{g1}=3.2$\,V, $V_\mathrm{g3}=1.15$\,V and $V_\mathrm{g5}=9.0$\,V. The bias voltage is $V_\mathrm{sd}=4$\,mV and therefore bias triangles appear larger in e) compared to b). The microwave power at the input of the feedline is $\sim-60$\,dBm for both measurements.}
%Applied voltages figures a) and b): $V_\mathrm{g1}=3.200$\,V, $V_\mathrm{g2}=1.282$\,V,
%$V_\mathrm{g3}=1.175$\,V,
%$V_\mathrm{g4}=1.758$\,V,
%$V_\mathrm{g5}=9.0$\,V.
%Applied voltages figures e) and f): $V_\mathrm{g1}=3.2$\,V, $V_\mathrm{g2}=0.5$\,V,
%$V_\mathrm{g3}=1.15$\,V,
%$V_\mathrm{g4}=1.0$\,V,
%$V_\mathrm{g5}=9.0$\,V.
\label{fig:twoconfigs}
%file of figures a,b: 290_20~1/G2vsG4_scan.dat"
%file of figures e,f: 171_20~1/G2vsG4_scan_zoom.dat
\end{figure}

\begin{figure*}[!htb]
\centering
\includegraphics[width=\textwidth]{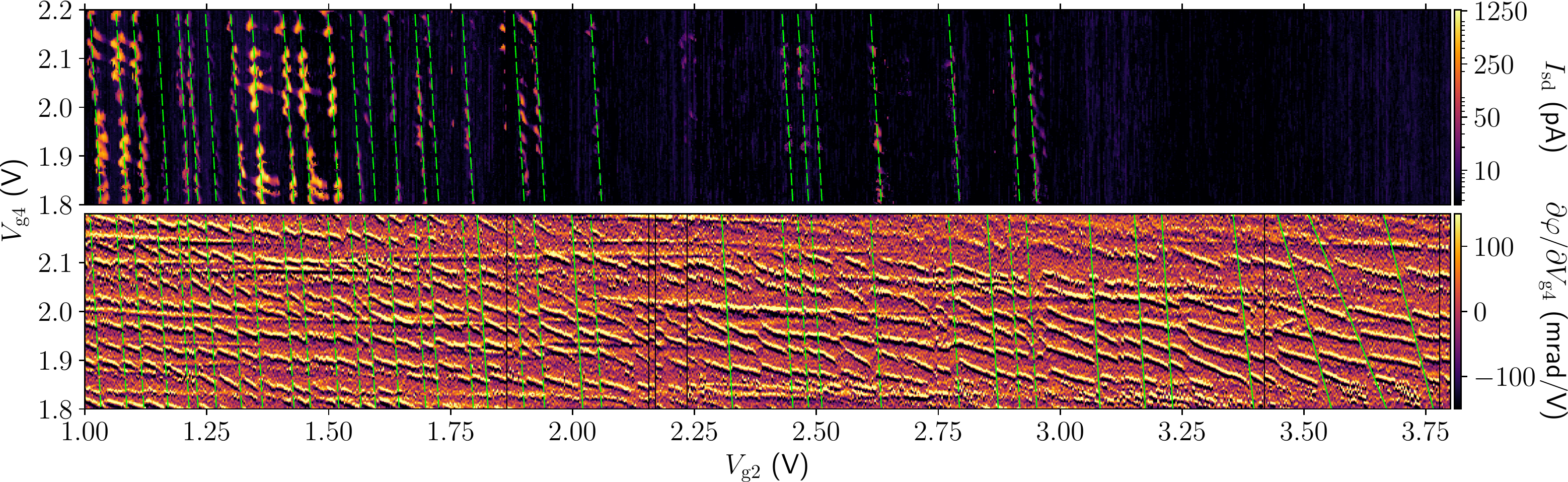}
\caption{Nanowire pinch-off. Top panel: map of dc current through the nanowire as a function of gate voltages $V_\mathrm{g2}$ and $V_\mathrm{g4}$, eventually vanishing completely above $V_\mathrm{g2}\approx 3$\,V as the nanowire is pinched off. The positions at which dot 1 is resonant with the lead are highlighted with green, dashed lines. Bottom panel: simultaneously measured resonator spectroscopy, $\partial \varphi/\partial V_\mathrm{g4}$. The resonator spectroscopy shows the same resonance conditions as in the top panel (green, dashed lines). However, additional transitions are observed (green, solid lines). Gate jumps are marked with vertical, black, solid lines. In this measurement, the other gate voltages are $V_\mathrm{g1}=3.2$\,V, $V_\mathrm{g3}=1.1$\,V and $V_\mathrm{g5}=9.0$\,V and the bias voltage is $V_\mathrm{sd}=2$\,mV and the readout frequency is $f_\mathrm{ro}=3.1259$\,GHz.}
\label{fig:nanowire_pinch_off}
\end{figure*}
\begin{figure}[tb]
\includegraphics[width=\linewidth]{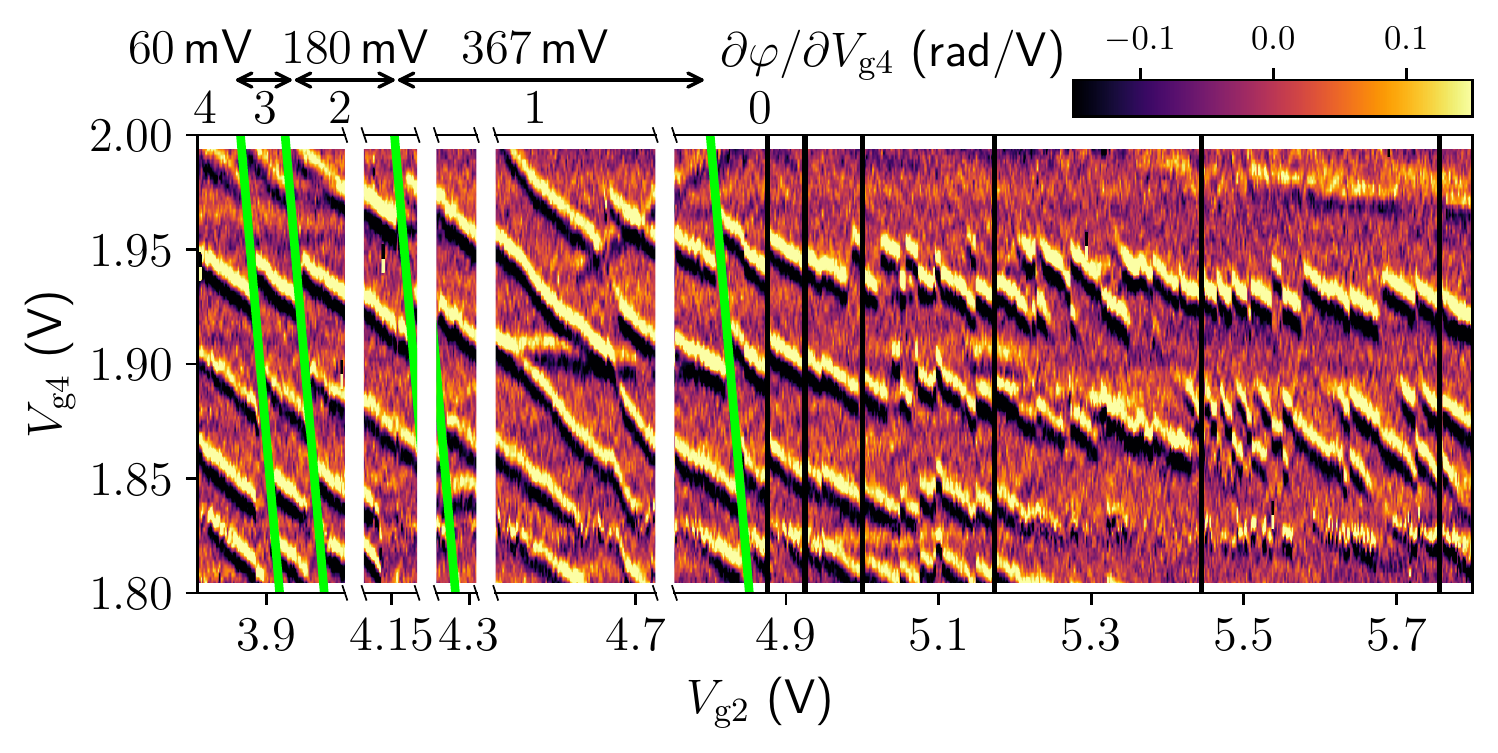}
\caption{Indications of last hole depletion.
Resonator spectroscopy, $\partial \varphi/\partial V_\mathrm{g4}$ as a function of gate voltages, $V_\mathrm{g2}$,  $V_\mathrm{g4}$.
Resonances correspond to right dot-right lead transition and characteristic discontinuities correspond to hole discharging from the left dot.
For gate voltages, $V_\mathrm{g2}>4.9$\,V, no further regular discontinuities are observed.
Instead, random jumps dominate the signal indicating that the last hole has been depleted.
The number of holes in the right dot is indicated by a number at the top of the graph.
Data repetition due to gate switchers has been omitted in the graph (see supplementary).
Here, the other gate voltages are $V_\mathrm{g1}=3.2$\,V, $V_\mathrm{g3}=1.1$\,V and $V_\mathrm{g5}=9.0$\,V.
The source drain bias voltage is $V_\mathrm{sd}=2$\,mV and the readout frequency is $f_\mathrm{ro}=3.1259$\,GHz.\label{fig:last_hole}}
\end{figure}

Due to the Fermi level pinning stemming from the staggered Si/Ge band-gap alignment, the Ge/Si core/shell nanowire is a hole conductor.
Therefore, by applying positive gate voltages, we define the barrier potentials on the gates $\mathrm{g}_1$, $\mathrm{g}_3$ and $\mathrm{g}_5$. This gives rise to the confinement potential of two quantum dots whose electrochemical potentials are tuned by the gates $\mathrm{g}_2$ and $\mathrm{g}_4$~\cite{Froning2018a}.

In the following, we investigate two different confinement configurations. The first configuration is schematically depicted in Figure~\ref{fig:twoconfigs}a). Here, two fairly symmetric quantum dots, the left (L) and the right (R), are formed between the gates g1 and g3 and between the gates g3 and g5. In this configuration, each dot couples to its respective neighbors as shown on the sketch in Figure~\ref{fig:twoconfigs}a).

In Figure~\ref{fig:twoconfigs}b), we plot a measurement of the direct current through the nanowire $I_\mathrm{sd}$ as a function of the voltages on gates g2 and g4. Because of Coulomb blockade, we measure a finite current through the nanowire only at the triple points at which the electrostatic potential of both dots is aligned with the electrostatic potential of the leads. By connecting the triple points (dashed white lines in Figure ~\ref{fig:twoconfigs}b)), we find the charge-stability diagram in the shape of a honeycomb pattern~\cite{van2002electron}.

Simultaneously to measuring the current through the nanowire, we send a microwave signal through the feedline at a frequency close to the resonance frequency $f_r$. We perform dispersive gate sensing by measuring the phase change of the transmitted signal and plot it in Figure~\ref{fig:twoconfigs}c) as a function of gate voltages.
As the resonator is capacitively coupled to the quantum dots via one of the plunger gates, it is sensitive to their effective admittance~\cite{delbecq2011coupling,Frey2012,ranjan2015clean}. Therefore, by sending a signal through the feedline at a frequency close to the resonator frequency, changes in the transmission amplitude and phase can be detected when the quantum dot admittance changes. Indeed, we note that in the plotted phase response, one can clearly identify the honeycomb pattern of the charge-stability diagram. Whenever the electrochemical potential between a dot and its respective lead, or between the two dots, is aligned, a shift in the phase response is observed. The charge-stability diagram that we gain from both dc and rf measurements are well described by a capacitance model~\cite{van2002electron}. By considering the change of the number of electrons when changing the gate potentials and using the source-drain bias triangles as an absolute energy scale, we fit the data according to the recipe described in Appendix A of Ref.~\cite{scarlino2021situ}. We extract the capacitances that are specified in Tab.~\ref{tab:caps}. 
\begin{table}[tb]
    \centering
    \begin{tabular}{c|c|c}
    & 2a,b,c) & 2d,e,f)\\
    \hline
         $C_\mathrm{g2,dL}$ (aF)&$3.4\pm 0.4$&$5.4\pm 0.8$\\
         $C_\mathrm{g4,dL}$ (aF)&$0.2\pm0.4$&$0.8\pm0.7$\\
         $C_{\Sigma,\mathrm{L}}$(aF)&$51\pm19$&$15\pm7$\\
         $C_\mathrm{g2,dR}$ (aF)&$0.4\pm0.4$&$0.1\pm0.6$\\
         $C_\mathrm{g4,dR}$ (aF)&$4.1\pm0.5$&$4.1\pm0.5$\\ 
         $C_{\Sigma,\mathrm{R}}$ (aF)&$57\pm20$&$20\pm12$\\
         $C_\mathrm{M}$ (aF)&$17\pm8$&$8\pm5$\\             
         \end{tabular}
    \caption{Gate-to-dot capacitances, where $C_{\mathrm{g}i,\mathrm{d}j}$ is the capacitance between gate gi and dot j ($i\in \{2,4\}$ and $j\in\{\mathrm{L},\mathrm{R}\}$. $C_{\Sigma,j}$ denotes the total capacitance of dot $j$ and $C_\mathrm{M}$ is the dot's mutual capacitance.}
    \label{tab:caps}
\end{table}

After, having demonstrated the possibility of detecting the charge-stability diagram by means of resonator spectroscopy, we tune the double quantum dot system into the configuration which is schematically depicted in Figure~\ref{fig:twoconfigs}f). The main difference to the previous configuration is the larger voltage on the gate g2, while the barrier gate voltages $V_\mathrm{g1}$ and $V_\mathrm{g3}$ are not changed significantly. This corresponds to a geometrically smaller dot L with a lower number of holes. Hence, the tunneling rate $t_L$ between the source and dot L, as well as the inter-dot tunneling rate $t_M$ are reduced. In this configuration, it is therefore not possible to measure these transitions using resonator spectroscopy. However, since $V_\mathrm{g5} =9.0\,V$ in both configurations, the remaining tunnel rate $t_R$ is, in first order, not affected, enabling us to use the dot R as a sensor for tracking Coulomb resonances of dot L~\cite{hutin2019gate,ansaloni2020single,chanrion2020charge,borjans2021spin}. 
When we progressively deplete dot L, the tunneling rate between the sensor dot and the drain always remains similar to the resonator frequency. This allows us us to track discharging lines of dot L despite the fact that the tunneling involving dot L happens at much lower frequencies and can therefore not directly be detected by dispersive resonator sensing.
%As the resonator is sensitive to the dot-lead resonances, we use it in order to form a charge sensor. For doing so, we tune the double-quantum dot into a different, more asymmetric configuration ($V_\mathrm{g1}=3.2$\,V, $V_\mathrm{g3}=1.1$\,V, $V_\mathrm{g5}=9.0$\,V). In this configuration, we use one dot as an ancilla or sensor dot that is capacitively coupled to a the dot under investigation.

Figure~\ref{fig:twoconfigs}d) shows the current through the nanowire in this configuration. We are still able to identify the locations of the triple points in the conductance measurement and calculate the capacitances as given in Table~\ref{tab:caps}.
Comparison of the conductance with the phase response in Figure~\ref{fig:twoconfigs}e) shows that the transmission through the feedline clearly exhibits a change in phase whenever the electrochemical potential of the sensor R is resonant with the one in the drain. We note characteristic jumps in the observed resonances.
These jumps correspond to discharging of a hole in the dot L. Therefore, by interconnecting jumps (green lines in Figure~\ref{fig:twoconfigs}f)), we determine the Coulomb resonances of the dot L.
 
The top panel of Figure~\ref{fig:nanowire_pinch_off} shows the current through the nanowire in a large range of $V_\mathrm{g2}$ in the same configuration as Figure~\ref{fig:twoconfigs}f). Coulomb resonances of the dot L that are observable in the current are highlighted by dashed, green lines. We note that when considering only the current, the largest gate voltage, at which a Coulomb resonance of dot L is observed, is $V_\mathrm{g2}\lessapprox 3$\,V. When examining the simultaneously measured resonator response in the bottom panel of Figure~\ref{fig:nanowire_pinch_off}, we identify several transitions that correspond to the sensor being in resonance with the drain. Here, for better visibility, we plot the derivative of the phase response with respect to the gate voltage $V_\mathrm{g4}$.
These sloped lines have kinks upon removing a hole from dot L because of the dots mutual capacitance. Therefore, by interconnecting the kinks, a Coulomb resonance of dot L is found. We identify several more Coulomb resonances of the dot L than in the dc measurement.
Note that the observed Coulomb resonances, highlighted by solid green lines, have a finite slope of $m=\Delta V_\mathrm{g4}/\Delta V_\mathrm{g2}\approx-18$ because of a finite capacitance between gate g4 and dot L.
The slope corresponding to these transitions changes for voltages $V_\mathrm{g2}\gtrsim3.4\,V$. This might be related to an (imperfect) potential landscape that makes the dot move to another equilibrium position below a certain number of holes. The slope corresponding to the last transition is $m=\Delta V_\mathrm{g4}/\Delta V_\mathrm{g2}\approx-3.9$ and remains the same subsequently, as shown in Figure~\ref{fig:last_hole}.

Inadvertent charge switching events occurring during this measurement can be rather easily identified because they happen suddenly, at a time scale smaller than the acquisition time of a single data point.
Such a single event appears as a (vertical) jump in gate voltage shifting the data along the entire axis, which we refer to as a gate jump from here on. 
Some of these gate jumps are indicated by vertical, black lines in the figures (e.g. around $V_\mathrm{g2}\approx 2.2$\,V in Fig.~\ref{fig:nanowire_pinch_off}).
Even for gate voltages $V_\mathrm{g2}$ much larger than the nanowire pinch-off current at 3\,V, several Coulomb resonances are found which cannot be identified when only considering the current through the nanowire.
We note that in the lower panel of Fig~\ref{fig:nanowire_pinch_off}, several horizontal features without any kinks are visible.
These are interpreted to originate from impurities coupling to the resonator, independent of the quantum dots.

Finally, with the goal in mind to deplete the last hole from dot L, we tune the gates into a third configuration in which we increase $V_\mathrm{g1}$ from  3.1\,V to 5.8\,V.
In this configuration, the nanowire is fully pinched-off and a direct current cannot be measured.
In Figure~\ref{fig:last_hole}, we plot the derivative of the phase on the resonator signal with respect to the gate voltage $V_\mathrm{g4}$ as a function of $V_\mathrm{g2}$ and $V_\mathrm{g4}$.
Once again, we identify resonances corresponding to tunnelling between dot R and the drain.
When connecting the characteristic shifts of these resonances, we obtain the parallel discharging lines (solid, green lines in Figure~\ref{fig:last_hole}) of the dot L. 
%with a slope of $m=\Delta_{V_\mathrm{g4}}/\Delta_{V_\mathrm{g2}}\approx-3.9$ which is consistent with the value obtained in Figure 2.
%The smaller slope as compared to Fig.~\ref{fig:nanowire_pinch_off} corresponds to a larger relative effect of gate g4 compared to gate g2 which makes sense as the large gate voltage of g2 pushes the dot away.
%The transition to smaller slopes is also clearly visible in Fig.~\ref{fig:nanowire_pinch_off} for voltages $V_\mathrm{g2}\gtrsim3.4$\,V where the current is already suppressed.

Since we work with larger gate voltages and thus a decreasing number of charges present in the wire, there is less screening and the wire becomes less stable, suffering from several gate jumps.
These gate jumps result in shifts along the $V_\mathrm{g2}$-axis towards less positive voltages.
In order to focus on the physics that corresponds to discharging of the dot L, those shifts are removed in Figure~\ref{fig:last_hole} where the removed regions are also clearly marked.
For completeness, the full data set can be found in Fig.~\ref{fig:extradatafig4} in the appendix. In Figure~\ref{fig:last_hole}, we observe a total of four sloped, parallel lines; each corresponding to discharging of a single hole from the dot L.
The last charging line is found at $V_\mathrm{g2}= 4.90$\,V, showing the position of the 1 to 0 hole transition in dot L, while the penultimate charging line is observed at $V_\mathrm{g2}=4.2$\,V (bottom axis) indicating the 2 to 1 hole transition.
We note that even after subtracting the additional voltage range, because of shifts along the $V_\mathrm{g2}$-axis due to gate jumps, the effective voltage distance between these two lines is $\Delta V_\mathrm{g2}\approx370$\,mV, much larger than the distance between any two previous discharging lines. 

For voltages larger than $V_\mathrm{g2}=4.80$\,V, beyond the last observed discharging line, the amount of gate jumps increases drastically.
They randomly shift the observed resonances in the gate-gate map and yield vertical disruptions of dot-lead resonances, even within a single vertical gate sweep (fast scan axis).
We therefore conclude that they correspond to the random charging and discharging of unwanted charge traps in proximity to the nanowire.
The absence of any further dot discharging lines appearing with a slope can give some confidence that indeed, the last hole was depleted from the left dot.
%However, one should be aware that the resonance locations on the $V_{g2}$ axis close to the end of the graph ($V_{g2}\approx5.8\,V$) are similar to the ones right after the last transition ($V_{g2}\approx4.9\,V$). Hence, despite we applied ~1\,V on gate 2 after the last transition, the charges resulting of charge traps might screen the effect of the gate (that, consequently, wouldn't be depleting the dot further). T.T
%It can indicate that the dot is still experiencing the same electric field (the effect of the gate would be counter by the one of the spurious transitions), despite we applied ~1\,V on gate 2 after the last transition.
We speculate that after depletion of the last hole from the dot, the sensor is more susceptible to unwanted charge traps as the screening by dot L vanishes.
Hence, the increase of random gate jumps is consistent with the interpretation that the discharging line at $V_\mathrm{g2}=4.80$\,V may correspond to discharging of the last hole.

To conclude, we demonstrate charge sensing of a Ge/Si core/shell nanowire double quantum dot by using a superconducting, high-impedance, on-chip NbTiN resonator.
Using bottom gates, we are able to define a double quantum dot in the nanowire and consistently map the characteristic charge-stability diagram by both direct current measurements and resonator spectroscopy.

By changing the electrostatic potentials on the gates, we tune the double quantum dot into a regime of a more isolated dot and a second, sensor dot which together with the resonator, we employ as a charge sensor of the first dot.
By increasing the gate voltages, we consecutively deplete holes from the dot.
We find that even in the regime where no current through the nanowire could be detected, because it is pinched-off, the sensor reveals several more hole discharging events while increasing the gate voltages.
Finally, we find indications of the depletion of the last hole from the nanowire.
Our measurements confirm that observing only the direct current through these type of nanowires is not a sufficient criterion for counting the absolute number of holes present in a quantum dot in Ge/Si core/shell nanowires.

The circuit-quantum electrodynamics architecture presented in this manuscript lays the foundations for realizing coherent charge-photon and spin-photon coupling based on semiconductor nanowires.
We expect that a reduction of the gate-gate and resonator-feedline capacitances will result in resonator quality factors by an order of magnitude larger.
Frequent gate jumps inhibited using the device as a spin qubit.

However, the charge stability of the system might be improved in the future by working on the quality of the oxides and nanowires.
%Very recently, after having finalized the measurements presented here, a different group presented strong hole spin-photon coupling where the holes are confined in a silicon MOS device~\cite{yu2022strong}.
Because similar nanowires as used in this work have been employed as spin qubits~\cite{Froning2021}, we anticipate that the improvements on the resonator in combination with a more stable nanowire device will enable strong charge photon and coherent spin-photon coupling in Ge/Si core/shell nanowires based on intrinsic spin-orbit interaction, as recently achieved in other material platforms~\cite{yu2022strong,ungerer2022strong}.

\pagebreak

This research was supported by the Swiss Nanoscience Institute (SNI), the Swiss National Science Foundation through grant 192027 and 179024, and the NCCR Spin Qubits in Silicon (NCCR-Spin). We further acknowledge funding from the European Union’s Horizon 2020 research and innovation programme, specifically the FET-open project AndQC, agreement No 828948, the FET-open project TOPSQUAD, agreement No 847471, and the European Microkelvin Platform (EMP), agreement No 824109. We also acknowledge support through the Marie Skłodowska-Curie COFUND grant QUSTEC, grant agreement N° 847471, and the Georg H. Endress foundation.
\bibliographystyle{apsrev4-1} % Tell bibtex which bibliography style to use
 \bibliography{ref}
\appendix
\renewcommand\thefigure{A.\arabic{figure}} 
\setcounter{figure}{0}
\section*{Additional data}
\begin{figure}[b]
    \centering
    \includegraphics[width=\linewidth]{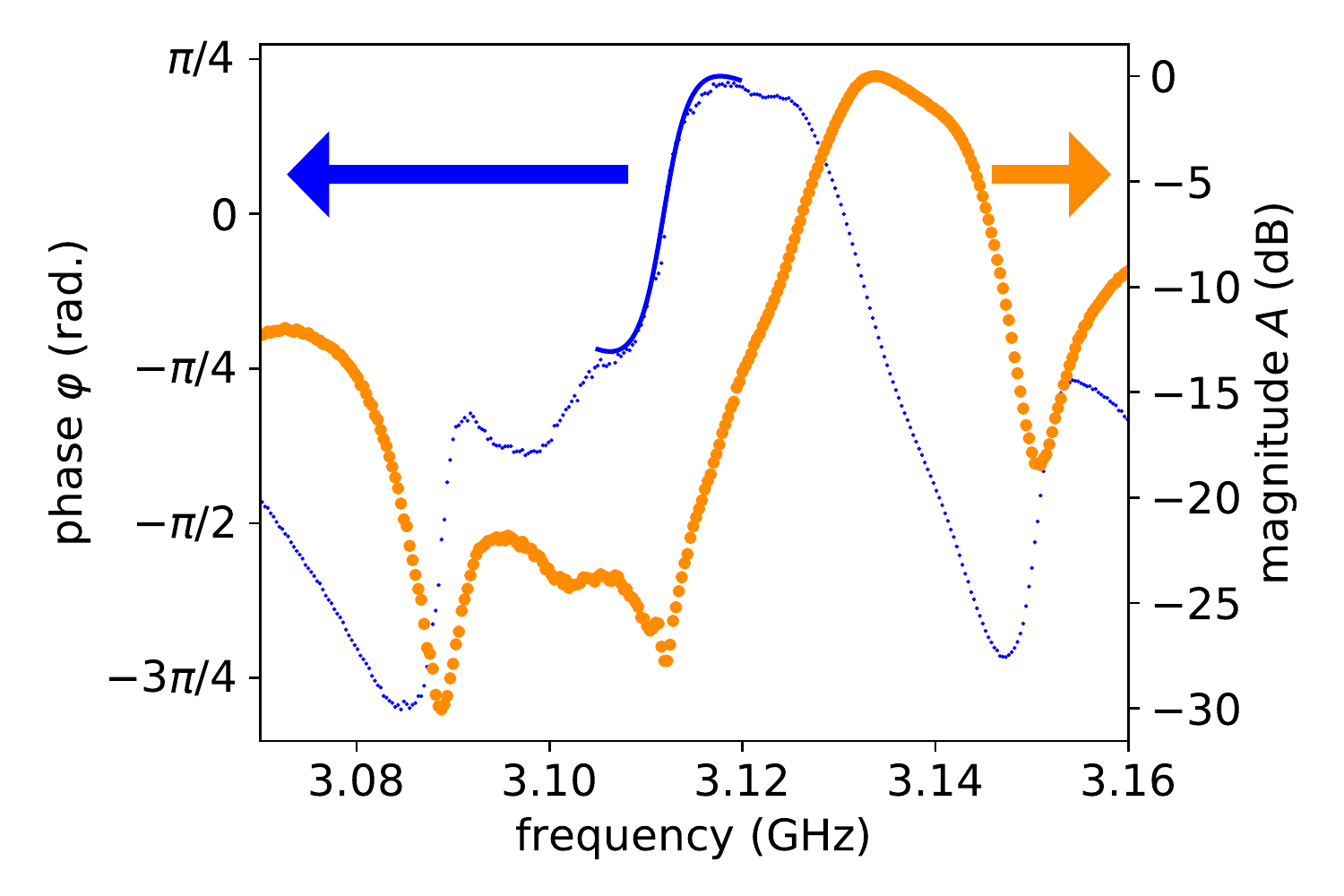}
    \caption{Transmission through the feedline in wide frequency range}
    \label{fig:extradatafig1}
\end{figure}
In Figure~\ref{fig:deviceoverview} in the main text, we show the resonance curve of the resonator. When looking at a wider spectral range, which is shown in Fig.~\ref{fig:extradatafig1}, it becomes apparent that the resonance is superimposed on a large standing wave background. Nonetheless, the resonator can be identified by considering a temperature-dependence scan, because its resonance frequency depends on the the large temperature-dependent kinetic inductance.

\begin{figure}[ht]
    \centering
    \includegraphics[width=\linewidth]{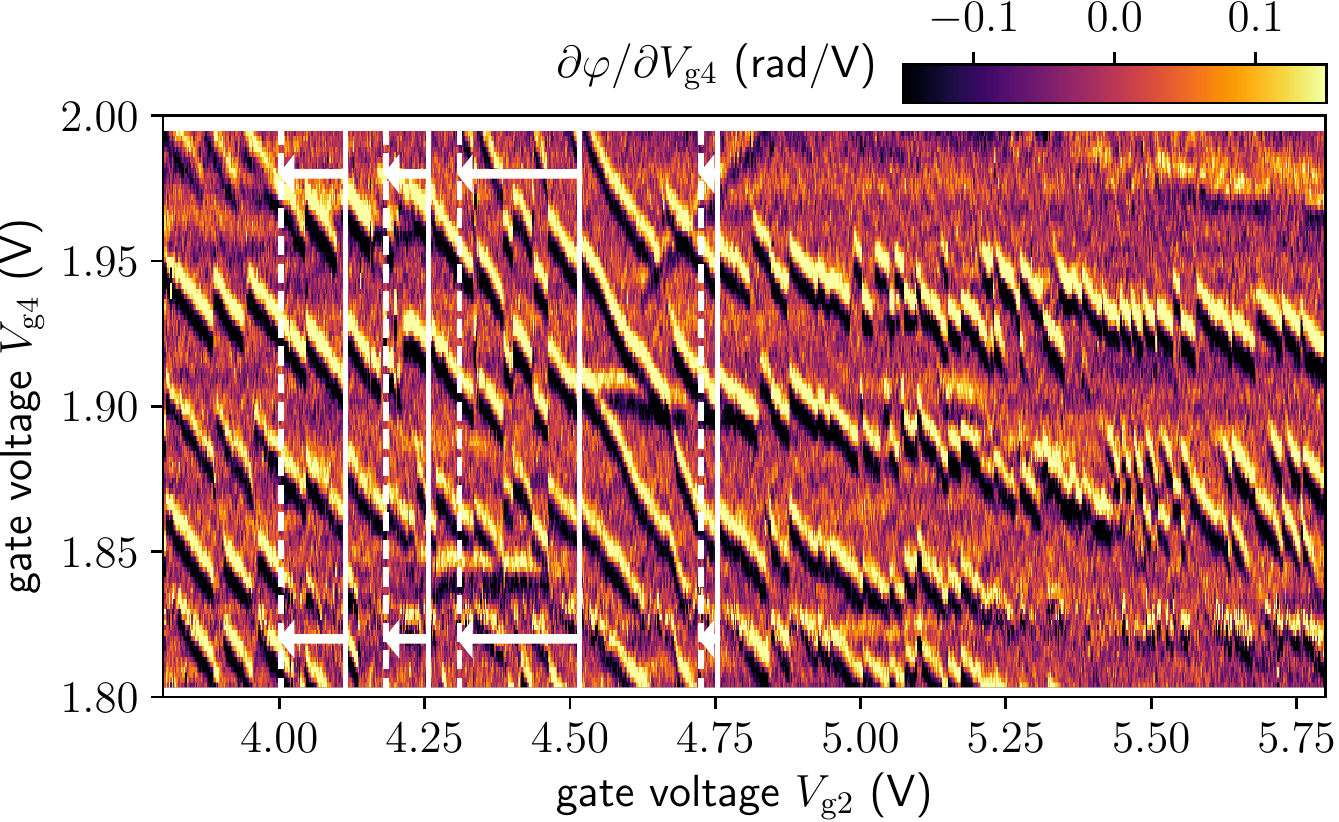}
    \caption{Resonator response as a function of gate voltage $V_\mathrm{g2}$ and $V_\mathrm{g4}$. This data set was used to create Fig.~\ref{fig:last_hole}. The solid, white lines show the positions of the gate jumps. In Fig.~\ref{fig:last_hole}, the data between the white, solid lines and the white, dashed lines, indicated by arrows, was omitted.}
    \label{fig:extradatafig4}
\end{figure}
During the measurement of the data presented in Fig.~\ref{fig:last_hole} in the main text, several gate jumps occurred. These gate jumps result in shifts along the $V_\mathrm{g2}$-axis. In order to focus on the relevant physics, we have omitted those shifts in Fig.~\ref{fig:last_hole}. Fig.~\ref{fig:extradatafig4} shows the complete data set where white annotations highlight which data was omitted in Fig.~\ref{fig:last_hole} (see caption of the figure).

%This is the begin of the appendix
%\begin{figure}
%\includegraphics[width=\linewidth]{FigA1/FigA2.pdf}
%\caption{phase of the transmission $S_{21}$ through the feedline exhibiting a resonance. The resonance was fitted to a theoretical curve (orange line) in order to extract the resonance frequency and Q factors.}
%\end{figure}
\end{document}